# Observation of Higher-Order Topological States in Acoustic Twisted Moiré Superlattice

Shi-Qiao Wu[1, *], Zhi-Kang Lin[1, *], Bin Jiang[1], Xiaoxi Zhou[1], Bo Hou[1], and Jian-Hua Jiang[1, †]

[1]*School of Physical Science and Technology & Collaborative Innovation Center of Suzhou Nano Science and Technology, Soochow University, 1 Shizi Street, Suzhou, 215006, China*

[*]*These authors contributed equally.*

[†]*Email: jianhuajiang@suda.edu.cn*

**Abstract:** Twisted moiré superlattices (TMSs) are fascinating materials with exotic physical properties. Despite tremendous studies on electronic, photonic and phononic TMSs, it has never been witnessed that TMSs can exhibit higher-order band topology. Here, we report on the experimental observation of higher-order topological states in acoustic TMSs. By introducing moiré twisting in bilayer honeycomb lattices of coupled acoustic resonators, we find a regime with designed interlayer couplings where a sizable band gap with higher-order topology emerges. This higher-order topological phase host unique topological edge and corner states, which can be understood via the Wannier centers of the acoustic Bloch bands below the band gap. We confirm experimentally the higher-order band topology by characterizing the edge and corner states using acoustic pump-probe measurements. With complementary theory and experiments, our study opens a pathway toward band topology in TMSs.

*Introduction.* TMSs are superlattices formed by two (or more) identical 2D periodic structures with a relative twisting angle [1]. The past years have witnessed TMSs as extraordinary materials of which the physical properties can be significantly tuned via the twisting angle, yielding various emergent phenomena, such as Mott insulators [2, 3], unconventional superconductivity [4-6], excitonic effects [7-9], and correlated Chern insulators [10], to name but only a few. Recently, the study of TMSs starts to extend from electronic systems to photonic [11-15] and phononic [16-18] systems, leading to the discovery of localization-delocalization transitions [11], bulk dispersion transitions in polaritons [12-15], and controllable phononic dispersions [16-18], etc. In photonic TMSs, the study of topological properties was focused on bulk dispersions of polaritons, yielding transitions between elliptic and hyperbolic isofrequency contours [12-15]. In contrast, the topological bulk-boundary responses in photonic and phononic TMSs, which form an important research direction, remains unexplored.

Here, we report on the theoretical design and experimental observation of higher-order topological states [19-24] in acoustic TMSs. Higher-order topological insulators in TMSs were initially proposed in electronic systems, leading to topological corner states and fractional corner charges [25, 26], which, however, have not yet been confirmed in experiments. To date, though higher-order topology has been studied in various models [27-35] and experimental systems [36-52] (see Ref. [53] for a review), it has never been realized in TMSs. The higher-order topological phase realized and observed in this work is the first of such a kind where the edge and corner states emerge due to the bulk band topology. Furthermore, we find that both the moiré twisting and the designed interlayer couplings play important roles in triggering the unique higher-order topology in the TMSs. Symmetry analysis and Wannier representations reveal unique topological invariants and abnormal corner states in the acoustic TMS, which are then confirmed in the acoustic pump-probe experiments.

*Acoustic TMS.* We construct the acoustic TMS from two identical layers. Each layer is a 2D honeycomb lattice of coupled acoustic resonators with a lattice constant $a = 56$mm [Fig. 1(a)]. The resonators are cylindrical and have the same geometry, i.e., the radius is $R = 0.2a$ and the height is $H_1 = 0.5a$. Within each layer, the nearest-neighbor couplings are realized by two parallel tubes of the diameter $d_1 = 0.08a$. In this way, each layer forms an acoustic analog of graphene of which the acoustic band structure is given in Fig. 1(b). One can see that a pair of Dirac points emerge at the $K$ and $K'$ points around 1.4kHz.

For simplicity and feasibility, we consider the TMSs with the moiré twisting angle $\theta = 21.78°$, i.e., the largest commensurate twisting angle. The interlayer couplings are realized by tubes of the diameter $d_2 = 0.3a$, which are only for the nearest interlayer couplings between A and B sites (denoted as the AB-type couplings; see Fig. 1(c) for illustration). In our acoustic TMSs, there is no direct interlayer couplings between the sites of the same type, i.e., there is no AA- or BB-type interlayer coupling. This design is inspired by the model of twisted bilayer graphene studied in Ref. [54], but it yields different physics here in acoustic systems.

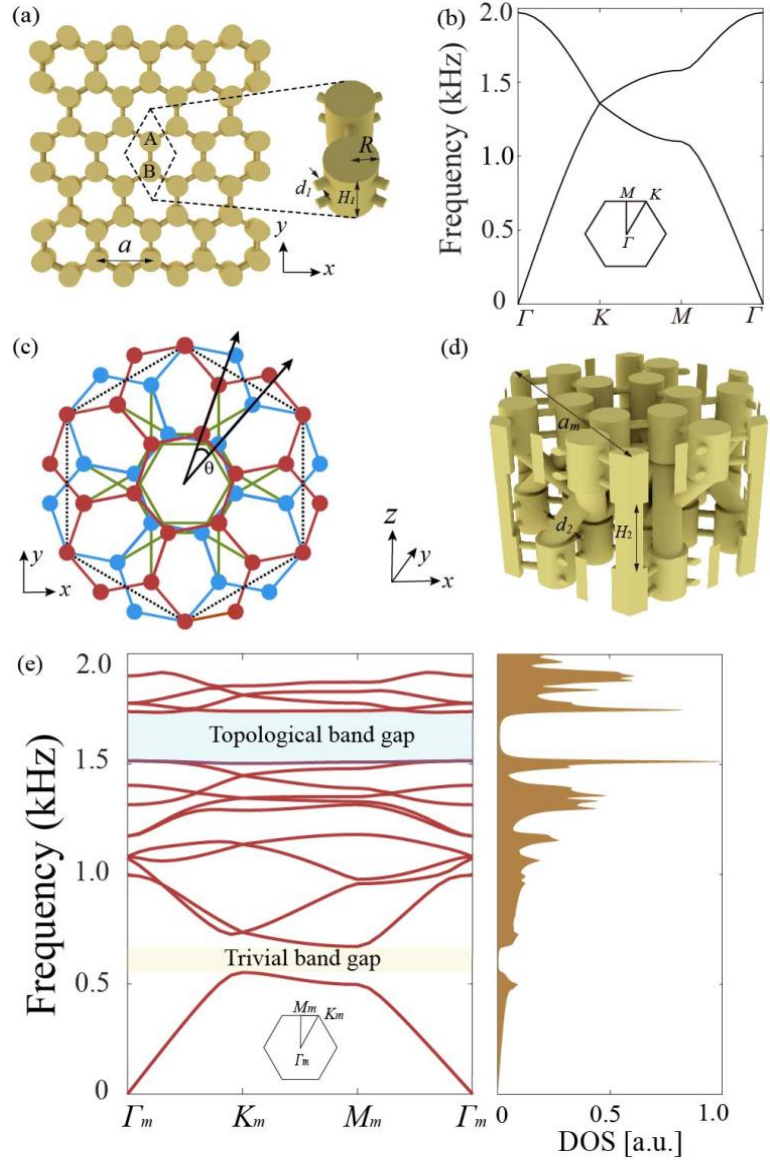

FIG. 1. Acoustic TMS. (a) Acoustic graphene realized by the honeycomb lattice of coupled acoustic resonators with A- and B-type sites. (b) Acoustic band structure of the lattice system in (a). (c) Schematic illustration of the acoustic TMS. Red represents the upper layer, while blue represents the lower layer. Dots represent the acoustic resonators, while the lines represent the tubes connecting them. Green lines stand for the tubes realizing the interlayer couplings. Black dashed lines depict the moiré unit-cell boundary. (d) A unit-cell of the acoustic model that realizes the TMS illustrated in (c). (e) Acoustic band structure of the TMS in (d). All geometry

parameters are specified in the main text. The density of states (DOS) of phonons are shown in the right panel. A Lorentzian broadening of 5Hz is used in the calculation of the DOS.

The acoustic structure of one moiré unit-cell is shown in Fig. 1(d) which has a lattice constant $a_m = \frac{a}{2\sin(\theta/2)}$. The two layers are separated by a surface-to-surface distance of $H_2 = 40$mm. The acoustic TMS has the $D_6$ point group symmetry and the time reversal symmetry. For the large twisting angle considered here, the magic angle physics for flat bands and related effects are irrelevant. In contrast, the strong interlayer couplings are crucial for the opening of the topological band gap, as elaborated below.

*Acoustic bands and higher-order topology.* The acoustic bands of the designed acoustic TMS are calculated and presented in Fig. 1(e). It is shown that there are two acoustic band gaps. The first band gap ranges from 553Hz to 670Hz, while the second band gap goes from 1520Hz to 1740Hz. Interestingly, the flat bands near the second band gap make the density of states of phonons very large near the valence and conduction band edges. Symmetry analysis of the Bloch states indicates that the first band gap is trivial, whereas the second band gap is a higher-order topological band gap (see discussions in the next section).

To understand the origin of the topological band gap, we present in Figure 2 the evolution of the acoustic bands with the interlayer couplings which are characterized by the diameter of the tubes connecting the A and B sites in different layers, $d_2$. We start from the limit with vanishing interlayer coupling in Fig. 2(a) (i.e., $d_2 = 0$) where the moiré twisting induces only the band folding effect due to the enlarged unit-cell. In this limit, the Dirac points (indicated by the arrow) are folded to the $K_m$ and $K'_m$ points (Throughout this Letter, the subscript $m$ denotes the quantities for the TMSs). There are 14 honeycomb unit-cells (28 lattice sites) within a moiré unit-cell. Therefore, there are 14 acoustic bands below the Dirac points. As the interlayer couplings increase, the Dirac points open a small gap [Figs. 2(b) and 2(c)] which must be due to the inter-valley scattering since there is no time-reversal breaking mechanism here [54]. Further increase of the interlayer couplings, however, closes the Dirac gap and opens another band gap below the Dirac points, i.e., between the eleventh and the twelfth bands [Fig. 2(d)]. The latter band gap increases with increasing interlayer couplings, reaching to a large band gap with higher-order topology at $d_2 = 0.3a$ [Figs. 2(e)-2(f)].

Our analysis indicates that the above scenario and the higher-order topology depend on the interlayer couplings and the symmetry of the TMSs. First, as shown in Supplemental Material [55], if the AB-type interlayer couplings are replaced by the AA- and BB-type interlayer couplings, the topological band gap opens instead between the eighth and ninth bands when $d_2 > 0.2a$. Second, the higher-order topology relies on the crystalline and time-reversal symmetries. The $D_6$ point group symmetry includes the $C_{6z}$ rotation symmetry around the $z$ axis and the $C_{2x}$ rotation symmetry around the $x$ axis. The latter transforms the upper layer into the lower layer. There are also the $C_{2y}$ rotation symmetry and other equivalent two-fold rotation symmetries that transform between the two layers (e.g., the $C_{2n}$ rotation symmetry around the 30° axis and the $C_{2m}$ rotation symmetry around the 120° axis in the x-y plane between the two layers).

The above symmetries are crucial for the higher-order band topology. For instance, the $C_{2z}$ symmetry guarantees a quantized Stiefel-Whitney number, $v_2$, due to an odd number of parity inversion between the M and $\Gamma$ point for all the bands below the band gap [25],

$$v_2 = (\frac{3}{2}\#M_2^{(2)} + \frac{1}{2}\#\Gamma_2^{(2)}) \bmod 2 \qquad (1)$$

Here, $\#M_2^{(2)}$ and $\#\Gamma_2^{(2)}$ denote the numbers of odd-parity bands below the gap at the $M$ and $\Gamma$ points, respectively. From the symmetry properties of the acoustic bands (see details in Supplemental Material [55]), we find $v_2 = 1$ which indicates nontrivial higher-order topology.

More completely, the band topology in crystalline materials with the $D_6$ point group and time-reversal symmetries is characterized by these symmetry indicators [56], $\chi_M = \#M_1^{(2)} - \#\Gamma_1^{(2)}$ and $\chi_K = \#K_1^{(3)} - \#\Gamma_1^{(3)}$. Here, $\#M_1^{(2)}$ and $\#\Gamma_1^{(2)}$ are the numbers of the even-parity Bloch bands below the band gap at the $M$ and $\Gamma$ points, respectively, while $\#K_1^{(3)}$ and $\#\Gamma_1^{(3)}$ denote the numbers of bands below the gap with the $C_{3z}$ eigenvalue 1 at the $K$ and $\Gamma$ points, respectively. From the symmetry properties of the acoustic bands, we find that $\chi_M = -2$ and $\chi_K = -2$. In fact, due to the conservation of the number of bands below the band gap, one can deduce that the Stiefel-Whitney number is related to $\chi_M$ as, $\nu_2 = \frac{\chi_M}{2}$ mod 2. Our analysis also indicates that the set of bands below the first and the second band gap exhibits fragile topology, yet all the bands below the second band gap in total are not fragile but Wannier representable, leading to higher-order topology with filling anomaly [56] (see Supplemental Material [55]).

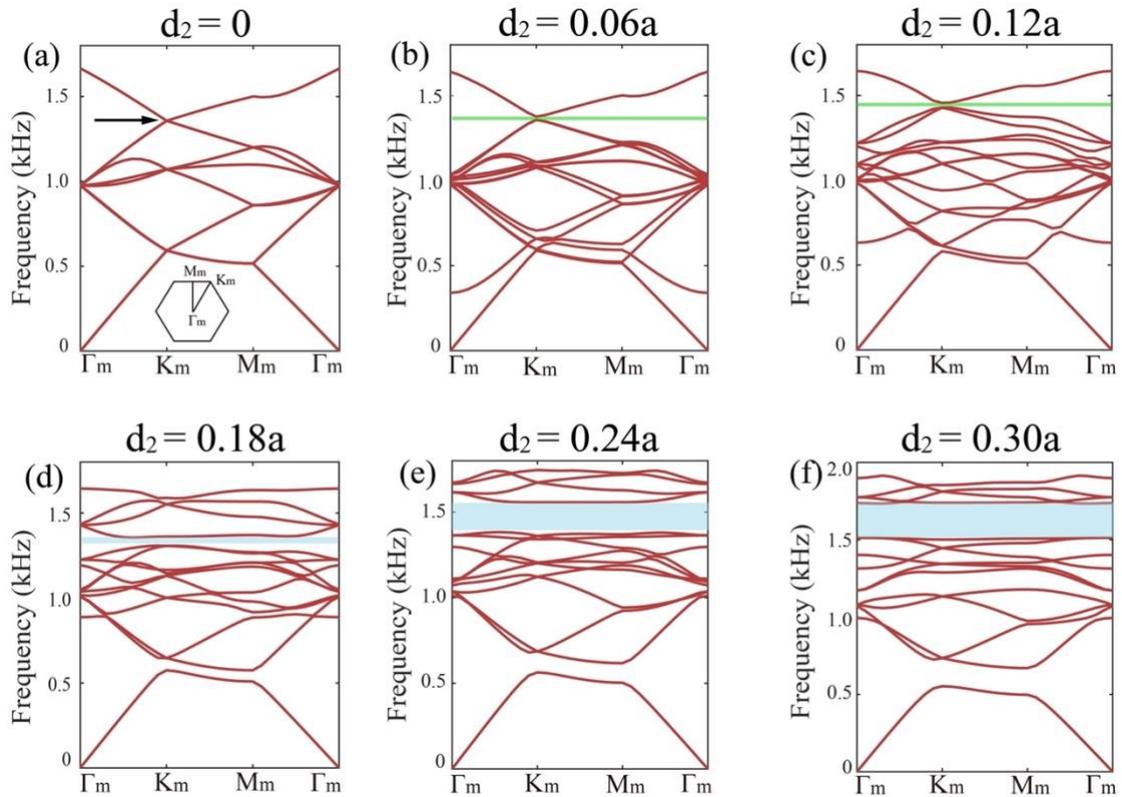

FIG. 2. Evolution of the acoustic band structure with the interlayer couplings. (a)-(f) Acoustic band structures with interlayer couplings tuned via the diameter $d_2$ of the tubes connecting the nearest A and B sites in different layers as depicted in Figs. 1(c)-1(d). The black arrow in (a) indicates the Dirac point. The green zones in (b) and (c) indicate the band gap opening at the Dirac points. The cyan zones in (d)-(f) indicate the band gap opening below the Dirac points.

*Calculated edge and corner states.* The above topological invariants readily lead to several important consequences. First, since the system is not in the fragile phase, the acoustic bands can be characterized by symmetric and localized Wannier functions. The nontrivial higher-order topology can be characterized by the Wannier centers of the acoustic bands which are not at the moiré unit-cell center. In finite systems, the Wannier centers at the edge and corner boundaries lead to the emergence of the edge and corner states as well as fractional corner charges [56].

We calculate the eigenstates spectrum of a finite system of rhombus shape with 5×5 moiré unit-cells. The results are presented in Fig. 3(a), indicating the emergence of the edge and corner states in the second band gap. As stated above, these edge and corner states can be understood via the Wannier centers of the bands below the gap. Due to the constraint of the crystalline symmetry, the Wannier centers can only reside at the high-symmetry points, i.e., the Wyckoff positions of the moiré unit-cell [see inset of Fig. 3(b)]. There are eleven bands below the gap and hence eleven Wannier centers. We find that there are three Wannier centers at the $3c$ Wyckoff positions (corresponding to the 1st, 4th and 5th bands) and two Wannier centers at the $2b$ Wyckoff positions (corresponding to the 2nd and 3rd bands) as well as six Wannier centers at the $1a$ Wyckoff positions (see Supplemental Material [55] for detailed analysis).

As known from the Su-Schrieffer-Heeger model, when the Wannier centers are exposed to the edge boundaries, they give rise to the edge states in the band gap [see Fig. 3(b)]. Meanwhile, there are Wannier centers at the four corner boundaries of the system, i.e., two obtuse-angle (type-I) corners and two acute-angle (type-II) corners. The type-I and type-II corners correspond to different geometries and thus support corner states with different frequencies. We find that there are two corner states at each corner in the bulk band gap which is related to the fact that each corner exhibits a two-fold rotation symmetry: A type-I corner has the $C_{2m}$ symmetry, while a type-II corner has the $C_{2n}$ symmetry [see Fig. 3(b)]. At each corner boundary, one corner state is even under the two-fold rotation, while the other corner state is odd under the same rotation. We find that the odd corner states have lower frequencies than the edge states, whereas the even corner states have higher frequencies than the edge states. Meanwhile, the even (odd) corner states at the type-II corners have lower frequencies than the even (odd) corner states at the type-I corners. In addition to these corner states in the bulk band gap, there are corner states in the bulk continua which form bound states in continuum. The total number of corner states within and outside the bulk band gap are consistent with the number of Wannier centers exposed to the corner boundaries (see Supplemental Material [55] for details).

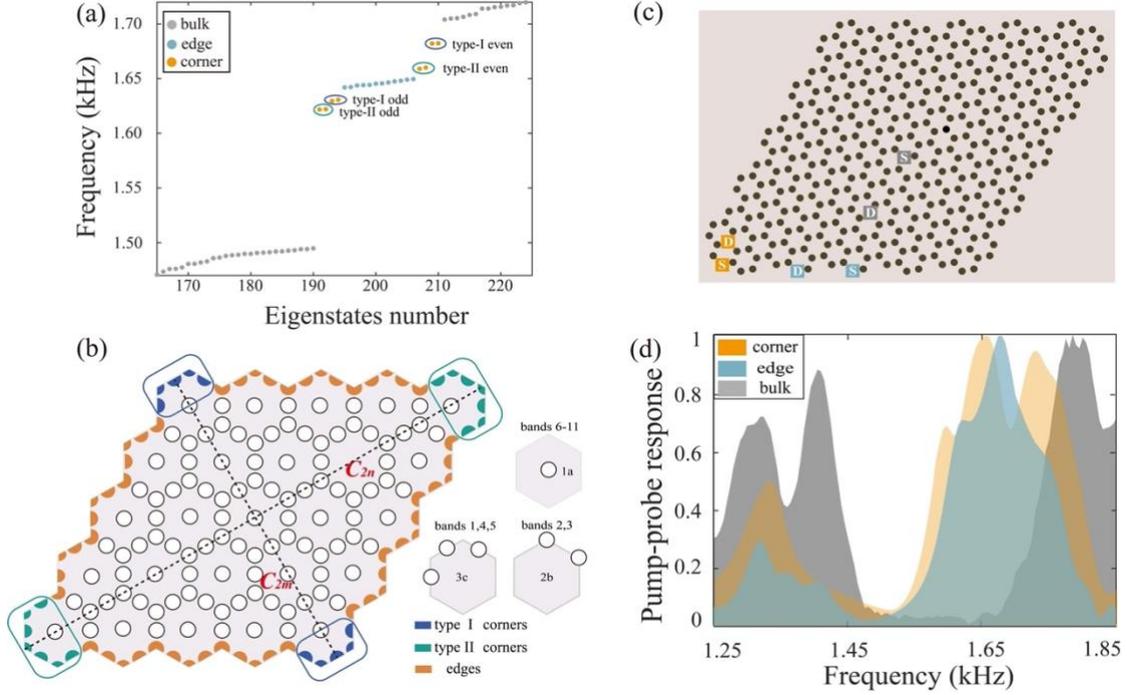

FIG. 3. Higher-order topology in the acoustic TMS. (a) Calculated eigenstates spectrum for a finite rhomboid structure with 5×5 moiré unit-cells. (b) Wannier centers of the bands below the second band gap for the structure in (a). (c) Schematic of the pump-probe measurements. The orange, bluish-gray and gray squares denote the pump-probe configurations for the corner, edge and bulk pump-probe responses. In experiments, the source (detector) is realized by a small headphone (microphone) inserted into the resonator labeled by the square with the letter S (D) for each pump-probe configuration. (d) Measured acoustic pump-probe responses for the corner, edge and bulk configurations. The frequency of the source varies from 1.25kHz to 1.85kHz, covering the second band gap.

*Acoustic pump-probe responses.* To verify experimentally the existence of the edge and corner states, we perform two kinds of measurements. First, we measure the acoustic pump-probe responses in the bulk, edge and corner regions to confirm the resonances within the bulk band gap at the edge and corner boundaries. We exploit three pump-probe configurations to study the spectroscopy in the bulk, edge and corner regions [see schematic illustration in Fig. 3(c)]. By varying the excitation frequency of the source (a small headphone), we measure the acoustic signal at the detector (a small microphone) in the frequency range from 1.25kHz to 1.85kHz. The amplitude of the measured signal gives the acoustic pump-probe responses shown in Fig. 3(d).

From Fig. 3(d), we find that the pump-probe response in the bulk region indicates a clear band gap between 1.5kHz and 1.7kHz, which is consistent with the calculated second band gap from 1520Hz to 1740Hz. Within this bulk spectral gap, there are resonance peaks in the measured pump-probe responses in the edge and corner regions. Fig. 3(d) shows that the corner pump-probe response has two main peaks, while the edge pump-probe response has one main peak between the two corner peaks. These features are coincident with the eigenstates spectrum in Fig. 3(b) where there are four corner states above the edge states and four corner states below the edge states. Such consistency indicates the emergence of the edge and corner states within the bulk topological gap.

*Measuring the eigenstates of a finite TMS.* We now use another kind of experimental signatures to confirm the edge and corner states induced by the higher-order topology. By exciting the bulk, edge and corner states at various frequencies in different regions, we can directly measure their wavefunctions through scanning the acoustic pressure field in the whole sample.

Figure 4(a) shows the fabricated sample that is used in the measurements. Figs. 4(b)-4(f) give the measured acoustic pressure profiles for various source locations and excitation frequencies. When the source is placed in the bulk region and set to an eigen-frequency in the bulk continua [e.g., 1420Hz in Fig. 4(b) and 1750Hz in Fig. 4(f)], the detected acoustic wave patterns show features consistent with the extended bulk states. In comparison, when the source is placed in the corner region and set to an eigen-frequency of the corner states [e.g., 1620Hz in Fig. 4(c) and 1660Hz in Fig. 4(e)], the detected acoustic pressure profiles manifest features of the localized corner states. These results confirm the observation of the bulk and corner states that are consistent with the calculations in Figs. 1 and 3. Similarly, the measurements confirm the features of the edge states when the source is placed in the edge region and set to an eigen-frequency of the edge states [e.g., 1645 Hz in Fig. 4(d)]. In Fig. 4, only five measured acoustic pressure profiles are shown, while more measured acoustic wave patterns are presented in Supplemental Material [55]. All the measurements confirm the edge and corner states emerging in the bulk band gap due to the higher-order topology, which is consistent with the calculations.

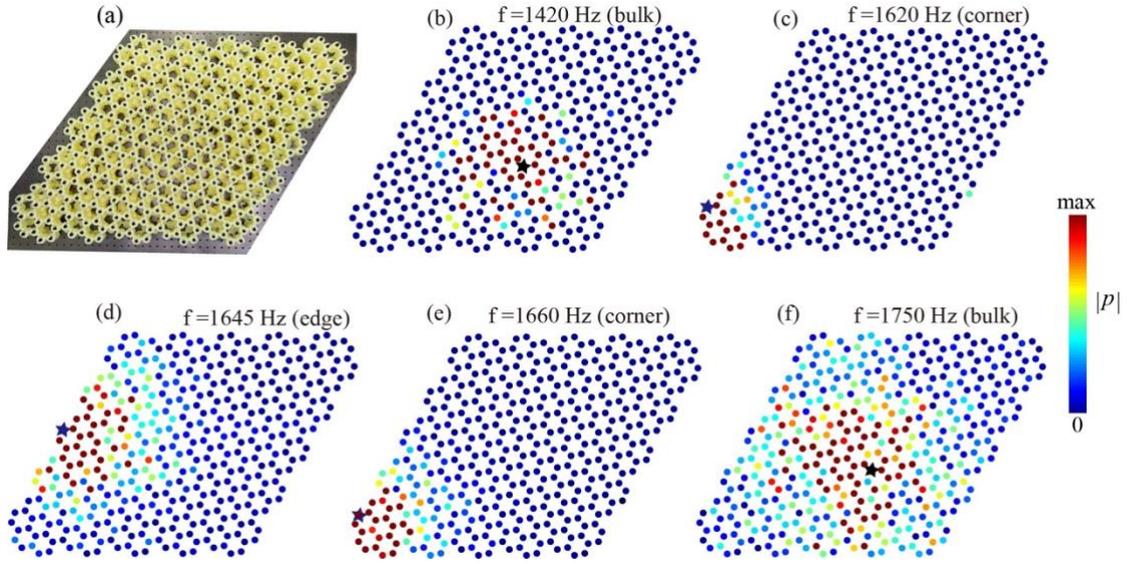

FIG. 4. Measured acoustic wavefunctions. (a) A photograph of the fabricated acoustic TMS with 5×5 moiré unit-cells. (b)-(f) Measured acoustic wavefunctions (i.e., the distributions of the acoustic pressure amplitude $|p|$) for five different excitation schemes. In (b) [(f)] the excitation frequency, 1420Hz (1750Hz), is in the bulk continuum below (above) the topological gap, while the source (labeled by the star) is in the bulk region. In (c) and (e) the excitation frequencies are close to the eigen-frequencies of the corner states below and above the edge states, respectively, while the source (labeled by the star) is in the corner region. In (d) the excitation frequency overlaps with the edge states, while the source is in the edge region.

*Conclusion and outlook.* The study of topological phenomena in TMSs is a frontier remains to be explored. Here, the observed higher-order topological phase in acoustic TMSs demonstrates the unique topological properties of TMSs and paves the way for future studies of topological physics in various TMSs.

*Note added.* — While our experimental work was completed and our manuscript was under preparation, a preprint was posted online which presents an experimental study on the photonic analog of twisted bilayer graphene and the higher-order topology [57].

*Acknowledgments*. This work was supported by the Natural Science Foundation of China (NSFC) (Grant Nos. 12074281, 12047541, and 12074279), the Major Program of Natural Science Research of Jiangsu Higher Education Institutions (Grant No. 18KJA140003), the Jiangsu specially appointed professor funding, and the Academic Program Development of Jiangsu Higher Education (PAPD). We thank Prof. Zhi Hong Hang for sharing his laboratory for part of the experimental measurements.